\documentclass[12pt]{article}
\usepackage{graphicx}
\usepackage{amsfonts}
\usepackage{amssymb}
\newcommand{\nin}{\noindent}
\newcommand{\be}{\begin{equation}}
\newcommand{\ee}{\end{equation}}
\newcommand{\bea}{\begin{eqnarray}}
\newcommand{\eea}{\end{eqnarray}}
\newcommand{\br}{\hskip .25cm/\hskip -.25cm}

\newcommand{\nn}{\nonumber\\}

\newcommand{\ol}{\overline}

\begin{document}
\begin{flushleft}
KCL-PH-TH/2011-6
\end{flushleft}
\vspace{0.5cm}
\begin{center}

{\bf{\Large Properties of a consistent Lorentz-violating Abelian gauge theory}}

\vspace{1cm}

{\bf J.~Alexandre}\footnote{jean.alexandre@kcl.ac.uk} and {\bf A.~Vergou}\footnote{ariadni.vergou@kcl.ac.uk}\\
King's College London, Department of Physics, London WC2R 2LS, UK

\vspace{1cm}

{\bf Abstract}

\end{center}

A Lorentz violating modification of massless $QED$ is proposed, with higher order space derivatives for the photon field. 
The fermion dynamical mass generation is studied with the Schwinger-Dyson approach. Perturbative properties of the model
are calculated at one-loop and discussed at higher-order loops, showing the consistency of the model. 
We explain that there is no contradiction with the definition of the speed of light $c$, although fermions
see an an effective light cone, with a maximum speed smaller than $c$.

\vspace{1cm}

\section{Introduction}

Recently, field theories with higher-order space derivatives have attracted attention, because of the improvement 
of graph convergence 
\cite{lifshitz}, at the price of violating Lorentz symmetry at high energies. 
Ghosts are not introduced by this procedure, since the time derivative order remains minimal, such that 
no new pole appears in the propagator of particles.
Also, theories based on anisotropic scaling of space and time (Lifshitz type theories) allow new 
renormalizable interactions, and for
example a renormalizable exponential potential, in 3+1 dimensions, has been studied in \cite{LL}. Finally, 
a renormalizable Lifshitz type theory of 
Gravity has been proposed, which could lead to Quantum Gravity \cite{Horava}.\\
A Lorentz-violating extension of massless Quantum Electrodynamics ($QED$) is discussed here, 
where higher order space derivatives are introduced for the photon field. 
This model has isotropic scaling in space time though, such that the higher order space derivatives involve a mass scale $M$, whose
role will be discussed a bit further. Higher order space derivatives were  already considered in a modified 
Dirac equation \cite{Pavlopoulos}, where the resulting phenomenology concerning gamma ray bursts is also discussed.
Higher order space derivatives could arise, for example, from quantum gravitational space time foam \cite{foam}.
The mass $M$ naively suppresses the effect of higher order derivatives in the IR,  
but as we will see, an IR signature of these higher orders remains: a fermion mass is generated dynamically. 
This mass, although proportional to $M$, is orders of magnitude smaller than $M$ 
since it is suppressed by an exponentially small function of the fine structure constant.
We will study here this mass generation, using the non-perturbative Schwinger-Dyson approach, and we will discuss 
the properties of the result.
We note here that another alternative to the Higgs model has been proposed in the context of anisotropic theories \cite{nohiggs},
and dynamical mass generation has been studied in \cite{fourfermion} for a Lifshitz-type
four-fermion model, and in \cite{yukawa} for a Lifshitz type Yukawa model.\\
We mention here the first works on dynamical mass generation in $QED$
\cite{johnson}, where a relation between the bare and dressed masses is derived, which involves a cut off $\Lambda$. In this
context, it was shown that, in order to have a finite theory, the limit $\Lambda\to\infty$ implies that the bare mass 
should vanish, such that the dressed mass must be of dynamical origin. The Schwinger Dyson equation 
with a finite cut off was studied later \cite{lambda}, where a critical value for the fine structure constant was found
in order to have dynamical mass generation in the limit $\Lambda\to\infty$. A summary of these results can be found in 
\cite{miransky}.\\
Once the fermion dynamical mass generation is demonstrated, we study the perturbative quantization of the model, taking
into account the fermion mass. 
We show that the model leads to the same renormalization flows as $QED$, when one considers the running scale $M$. The only difference is
the effective light cone seen by fermions, since space a time derivatives get different 
quantum corrections. 
We find that the maximum speed for fermions is smaller than the speed of light, by terms of one-loop order,
showing that the model is consistent with causality.
This is not in contradiction with the definition speed of light,
since the latter is obtained in the massless limit, which cannot be taken for fermions, due to dynamical mass generation.
The speed of light is given by the dispersion relation for the gauge field, which is not altered by the Lorentz-violating 
model proposed here. \\
We note finally that more general higher order derivative extensions of $QED$ have been  
presented in \cite{kostelecky} and references therein. 
In these works, the authors consider Lorentz-violating vacuum expectation values for tensor fields,
which allow the introduction of higher order derivatives for the photon field. They explain that the Lorentz-violating Lagrangians
can be written as the Lagrangian for $QED$ in a medium, and they study for example the corresponding birefringence effects of the vacuum.
Our present study corresponds to a specific case of the latter models, where we develop quantum properties of a given Lorentz-violating 
extension of $QED$.\\
In Section 2 we define the model and discuss some of its classical features. The fermion dynamical mass is then derived in 
Section 3, and the properties of the quantum theory are developed in Section 4, where we first focus on one-loop, and 
then develop general arguments for higher-order loops. The Conclusion presents future extensions involving non-Abelian 
gauge fields, and an alternative to the Higgs mechanism. Finally, the Appendix contains the details of one-loop 
calculations.

\section{Model}

The Lorentz-violating Lagrangian considered here is 
\be\label{bare}
{\cal L}=-\frac{1}{4}F^{\mu\nu}\left(1-\frac{\Delta}{M^2}\right)F_{\mu\nu}
-\frac{\xi}{2}\partial_\mu A^\mu\left(1-\frac{\Delta}{M^2}\right)\partial_\nu A^\nu 
+i\ol\psi\br D\psi,
\ee
where $D_\mu=\partial_\mu+ieA_\mu$, and $\Delta=-\partial_i\partial^i=\vec\partial\cdot\vec\partial$ 
(the metric used is (1,-1,-1,-1)),
which recovers $QED$ in a covariant gauge if $M\to\infty$. 
The Lorentz-violating terms have two roles: introduce a mass scale, necessary to generate a fermion mass, 
and lead to finite gap equation, as will be seen further. 
We stress here that $M$ regularizes only loops with an internal photon line, and that
another regularization is necessary to deal with fermion loops. 
Also, the Lorentz violating modifications proposed in the Lagrangian (\ref{bare}) do not 
alter the photon dispersion relation, which remains relativistic.\\
No higher order space derivatives are introduced for the fermions, for the following reason: 
in order to respect gauge invariance, such terms would need to be of the form
\be
\frac{1}{M^{n-1}}\ol\psi (i\vec D\cdot\vec\gamma)^n\psi~~~~~~n\ge 2,
\ee
such that new and non-renormalizable couplings would be introduced.\\

\vspace{0.5cm}

\nin{\it Equations of motion}\\
The classical equation of motion for the free gauge field is
\be
\left( 1-\frac{\Delta}{M^2}\right) \left( \Box\eta_{\mu\nu}-\partial_\mu\partial_\nu\right) A^\mu=0~,
\ee
and has two solutions. One is the usual plane wave, with the usual dispersion relation, and the other is solution of
\be\label{eqbis}
\left( 1-\frac{\Delta}{M^2}\right) A_\mu=0~.
\ee
If the solution depends on time and one spatial direction $x$ only, it reads
\be
A_\mu=K_\mu(t)e^{-Mx}~,
\ee 
where $K_\mu(t)$ is an homogeneous vector. If we consider 
a spherically symmetric solution, depending on time and the radial spatial coordinate $r$ only, we find
\be
A_\mu=K_\mu(t)\frac{e^{-Mr}}{r}~.
\ee
In both cases, the only regular solution occurs when $K_\mu=0$. As a consequence, the only physical solution of the classical equation
of motion for the free gauge field is the usual plane wave.\\

\vspace{0.5cm}

\nin{\it Lorentz transformation}\\
Finally, one can ask what the effect of a Lorentz transformation is on higher order space derivatives, as in the model (\ref{bare}). 
For this, we consider two coordinate systems
$(t,\vec r)$ and $(t',\vec r~')$, with the following transformation law
\bea\label{Lorentztransfo}
t&=&\gamma(t'+\vec v\cdot\vec r~')\\
\vec r&=&\vec r_\bot~'+\gamma(\vec r_\|~'+\vec v t')\nonumber~,
\eea
where the usual notations are used. In the Lorentz transformation (\ref{Lorentztransfo}), we have 
\be
-\partial_t^2+\Delta=-\partial_{t'}^2+\Delta'~~~~\mbox{with}~~~~
\partial_{t'}=\gamma\partial_t+\gamma\vec v\cdot\vec\nabla~,
\ee
such that 
\be\label{Delta'}
\Delta'=\Delta+(\gamma^2-1)\partial_t^2+2\gamma^2\vec v\cdot\vec\nabla\partial_t+\gamma^2(\vec v\cdot\vec\nabla)^2~.
\ee
Using eq.(\ref{Delta'}), the equation of motion $(M^2-\Delta')\phi(t',\vec r~')=0$ for a scalar field 
\be
\phi(t',\vec r~')=\phi(t,\vec r)=\phi_0\exp(i\omega t-i\vec k\cdot\vec r)~,
\ee 
leads then to the dispersion relation 
\be
\omega^2+2\omega\frac{\vec v\cdot\vec k}{v^2}+\frac{(\vec v\cdot\vec k)^2}{v^2}+\frac{M^2+k^2}{(v\gamma)^2}=0~,
\ee
such that
\be
\omega=-\frac{\vec v\cdot\vec k}{v^2}\pm\frac{i}{v\gamma}\sqrt{M^2+k^2-(\vec v\cdot\vec k)^2/v^2}~.
\ee
As a consequence, the corresponding solution is not observable, since it decays in a time of the order of the Plank time
(we do not take into account the non-physical solution increasing exponentially in time). 
Therefore the higher order time derivatives introduced by a Lorentz transformation do not affect the model presented here.

\section{Gap equation}

We review here the results of \cite{mdyn}. The Schwinger-Dyson equation for the fermion propagator is \cite{miransky}:
\be\label{SD}
G^{-1}-G_{bare}^{-1}=\int D_{\mu\nu}(e\gamma^\mu) G\Gamma^\nu,
\ee
where $\Gamma^\nu$, $G$ and $D_{\mu\nu}$ are respectively the dressed vertex, the dressed fermion propagator 
and the dressed photon propagator. This equation gives an exact self consistent relation between dressed 
$n$-point functions, and thus is not perturbative.
As a consequence, no redefinition of bare parameters can be done in order to absorb would-be divergences, 
and for this reason one needs this equation to be regularized by $M$. \\
In the studies of dynamical mass generation 
in $QED$ in the presence of an external magnetic field $B$ (``magnetic catalysis'' \cite{mag}), 
the gap equation, in the Lowest Landau Level approximation, is finite because of the mass scale $\sqrt{|eB|}$ 
- where $e$ is the electric charge - which plays the role of a gauge invariant cut off.
Another example of dynamical mass generation, however of a different nature, is the Debye screening for the photon, at 
finite temperature, which was found to be enhanced by a strong magnetic field \cite{Debye}.\\
In order to solve the Schwinger-Dyson equation (\ref{SD}), we consider the so-called ladder (or rainbow) approximation,
consisting in taking the bare vertex instead of $\Gamma^\nu$. It is known that this approximation 
is not gauge invariant \cite{miransky}, but, as we will see, 
the gauge coupling dependence of the dynamical mass is not affected by the choice of the gauge parameter $\xi$.
Then we will neglect corrections to the photon propagator, which is given by
\be\label{D}
D_{\mu\nu}^{bare}(\omega,\vec p)=\frac{i}{1+p^2/M^2}\left( \frac{\eta_{\mu\nu}}{-\omega^2+p^2}+
\zeta\frac{p_\mu p_\nu}{(\omega^2-p^2)^2}\right) ,
\ee
where $\zeta=1/\xi-1$, $p_0=\omega$ and $p^2=\vec p\cdot\vec p$. 
Also, we neglect the fermion wave function renormalization: 
only the fermion dynamical mass will be taken into account as a correction, such that the dressed fermion propagator 
will be taken as
\be\label{G}
G(\omega,\vec p)=i\frac{\omega\gamma^0-\vec p\cdot\vec\gamma+m_{dyn}}{\omega^2-p^2-m_{dyn}^2},
\ee
where $m_{dyn}$ is the fermion dynamical mass.
With these approximations, the Schwinger-Dyson equation (\ref{SD}) - involving a convergent integral - leads to 
\be\label{mdyn=}
m_{dyn}=\frac{\alpha}{\pi^2}\int_{-\infty}^\infty d\omega\int_0^\infty
\frac{p^2dp}{1+p^2/M^2}\frac{m_{dyn}(4+\zeta)}{(\omega^2+p^2)(\omega^2+p^2+m^2_{dyn})}~,
\ee
where $\alpha=e^2/4\pi$ is the fine structure constant.
This equation has the obvious solution $m_{dyn}=0$, and potentially a second solution, which must
satisfy the following gap equation, obtained after integration over the frequency $\omega$,
\be\label{intgap}
\frac{\pi}{(4+\zeta)\alpha}=\int_0^\infty\frac{xdx}{1+\mu^2x^2}\left( 1-\frac{x}{\sqrt{1+x^2}}\right),
\ee
where $\mu=m_{dyn}/M$ is the dimensionless dynamical mass, expected to be very small $\mu<<1$, and $x=p/m_{dyn}$.
Both terms in the last equation, if taken separately, lead to diverging integrals, with divergences which cancel each other.
An integration by parts for the second term leads then to
\bea\label{8}
\frac{2\pi}{(4+\zeta)\alpha}&=&\frac{1}{\mu^2}\int_0^\infty dx\frac{\ln(1+\mu^2x^2)}{(1+x^2)^{3/2}}\nn
&=&\frac{2}{\mu^2}\left( \ln\left(\frac{\mu}{2}\right)+\frac{\cosh^{-1}(1/\mu)}{\sqrt{1-\mu^2}}\right)\nn
&=&\ln\left( \frac{2}{\mu}\right) -\frac{1}{2}+{\cal O}(\mu^2\ln\mu)
\eea
and the fermion dynamical mass is finally given by
\be\label{mdyn}
m_{dyn}\simeq M\exp\left(-\frac{2\pi}{(4+\zeta)\alpha}\right).
\ee
Note that the expression (\ref{mdyn}) for $m_{dyn}$ is not analytic in $\alpha$, 
such that a perturbative expansion cannot lead to such a result, which
justifies the use of a non-perturbative approach. A perturbative expansion would lead to the solution $m_{dyn}=0$ only.\\
Among the two solutions $m_{dyn}=0$ and $m_{dyn}\ne0$, the physical system chooses the non-vanishing dynamical mass,
in order to avoid IR instabilities, not favorable energetically, which would otherwise occur in the theory.

\vspace{0.5cm}

\nin{\it Gauge dependence} \\
There is an obvious dependence on the gauge parameter $\zeta$, which has a consequence on the value of $m_{dyn}$, but 
the important point is the non-analytic $\alpha$-dependence of the dynamical mass, which is not affected by the choice of gauge: 
the resulting dynamical mass is of the form $M\exp(-c/\alpha)$, where $c$ is a constant of order 1. 
This feature is known in the studies of dynamical mass generation in $QED$ in the presence of an external magnetic field \cite{mag}.\\
It has been argued though, using the pinch technique \cite{pinch}, that for the calculation of physical quantities, which
must be gauge invariant, one should use the Feynman gauge $\zeta=0$ to calculate the fermion self energy. The reason for this is
the cancellation of terms arising from longitudinal contributions in the photon propagator. 
Nevertheless, if $M$ is the Plank mass, the Feynman gauge for the 
dynamical mass (\ref{mdyn}) leads to a too small result for the electron mass. But a multibrane scenario has been proposed in
\cite{branes}, where a Randal-Sundrum warp factor enhances the dynamical mass, which then acquires the right order of magnitude
for the electron, when $M$ is the Plank mass.

\section{Perturbative analysis of the model}

Taking into account the fermion mass $m_{dyn}$, 
we discuss here the perturbative quantization of the model, first at a one-loop, and we give then the general arguments
at higher order loops.

\subsection{One-loop properties}

{\it Fermion kinetic term}\\
We calculate in the Appendix the one-loop quantum corrections for the fermion kinetic term, which are different for time and
space derivatives. We find the corrections
\be
i\ol\psi \left((1+Z_0)\partial_0\gamma^0-(1+Z_1)\vec\partial\cdot\vec\gamma\right) \psi~, 
\ee
with
\bea
Z_0&=&-\frac{\alpha}{2\pi}\left( \ln\left( \frac{1}{\mu}\right) +4\ln2-2\right) +{\cal O}(\mu^2\ln(1/\mu)) \nn
Z_1&=&-\frac{\alpha}{2\pi}\left( \ln\left( \frac{1}{\mu}\right) +\frac{50}{9}-\frac{20}{3}\ln2\right)+{\cal O}(\mu^2\ln(1/\mu))~.
\eea
Note that the dominant term, proportional to $\ln(1/\mu)$, is the same for $Z_0$ and $Z_1$, since in the 
Lorentz symmetric situation ($M\to\infty$ for fixed fermion mass), we have $Z_0=Z_1$. 
Also, the coefficient $-\alpha/(2\pi)$ in front
of the dominant term $\ln(1/\mu)$ is the coefficient found in $QED$ in $4-\epsilon$ dimensions 
\be
Z_0^{QED}=Z_1^{QED}=-\frac{\alpha}{2\pi\epsilon}+\mbox{finite}~.
\ee  
An important remark should be made here: because of the result (\ref{mdyn}), the ratio $\mu$ is actually finite in the limit where
$M\to\infty$, and one could think that no counter term is necessary to absorb terms proportional to $\ln(1/\mu)$. 
But it is in fact necessary, in order 
to respect the loop structure. Indeed, if one keeps $\ln(1/\mu)$-terms in the renormalized theory, 
the would-be one-loop correction would become a tree-level one: 
\be
\alpha\ln\left( \frac{1}{\mu}\right) ={\cal O}(1)~.
\ee
Therefore, provided one treats $\ln(1/\mu)$ as a $1/\epsilon$ term in dimensional regularization, one gets the usual $QED$ one-loop 
structure.

\vspace{0.5cm}

\nin{\it Maximum speed for fermions}\\
After redefinition of the bare parameters in the minimal substraction scheme, where
only the term proportional to $\ln(1/\mu)$ is absorbed, the fermion dispersion relation is
\be
\left(1-\frac{\alpha}{\pi}[2\ln2-1]\right)^2\omega^2=\left(1-\frac{\alpha}{\pi}[25/9-(10/3)\ln2]\right)^2p^2+m_{dyn}^2~,
\ee
and the product of the fermion phase velocity $v_\phi$ and group velocity $v_g$ is then
\bea\label{phase}
v^2&\equiv&v_\phi v_g=\frac{\omega}{p}\frac{d\omega}{dp}\\
&=&\left( \frac{1-(\alpha/\pi)[25/9-(10/3)\ln2]}{1-(\alpha/\pi)[2\ln2-1]}\right)^2\nn
&=&1-\frac{2\alpha}{\pi}\left(\frac{34}{9}-\frac{16}{3}\ln2\right) +{\cal O}(\alpha^2)~<1~,
\eea
which shows that the effective light cone seen by fermions is consistent with causality. If we take $\alpha\simeq1/137$, we obtain
for the fermion maximum speed
\be\label{vpvg}
v\simeq1-1.9\times10^{-4}~.
\ee

\vspace{0.5cm}

\nin{\it Gauge invariance and speed of light}\\
We calculate in the Appendix the one-loop vertex for vanishing incoming momentum, and we find the corrections
\bea
\Gamma_{(1)}^0&=&Z_0e\gamma^0\nn
\Gamma_{(1)}^i&=&Z_1e\gamma^i~,
\eea 
such that gauge invariance is respected, since quantum corrections to the vertex are identical with corrections to the fermion 
kinetic term, for time and space components independently. 
As a consequence, as in usual $QED$, quantum corrections to the coupling constant are given by the polarization tensor, and arise
only from the wave function renormalization of the gauge field. Since the one-loop polarization tensor
doesn't include any internal photon line, the one-loop running of the coupling constant is therefore also the 
same as in $QED$.\\
Because $Z_0\ne Z_1$ and therefore the effective light cone seen by fermions involves the speed $v<1$, one 
might see a problem with the definition of speed of light. We argue here that it is not the case, because of dynamical 
mass generation for the fermion. Indeed, the speed of light $c$ is defined by
\be\label{limit}
c=\lim_{m\to0}~\frac{\omega}{|\vec p|}~,
\ee
for finite momentum $\vec p$ and frequency $\omega$.
But because the fermion is always massive, $m=m_{dyn}\ne 0$, the limit (\ref{limit}) cannot be taken, and the result (\ref{vpvg}) is
not in contradiction with the speed of light.
Such a conclusion was already obtained in \cite{yukawa} for a Lifshitz-type Yukawa interaction.\\
The speed of light is given by the gauge field dispersion relation, which is not modified at one-loop. 
We show in the next subsection that this conclusion still holds at higher order loops.

\subsection{Higher order loops}

\nin{\it Perturbative expansion}\\
We have seen at one-loop that, in order to recover the $QED$ renormalization structure, one needs to absorb terms proportional 
to $\ln(1/\mu)$ in the redefinition of bare parameters. If we assume that, after including the appropriate counter terms, 
we have a finite theory at $n-1$ loops, a graph calculated at $n$ loops will be of the form
\be\label{finite}
G^{(n)}=P^{(n)}\frac{M^\epsilon}{\epsilon}+Q^{(n)}\ln\left( \frac{1}{\mu}\right)+\mbox{finite}~,
\ee
where, compared to $(n-1)-$loop graphs:\\ 
$P^{(n)}$ contains graphs with an additional fermion loop, which needs to be regularized 
dimensionally;\\
$Q^{(n)}$ contains graphs with an additional photon line, which are regularized by $M$. \\
The $n$-loop counter term then needs to absorb terms proportional to both $1/\epsilon$ and $\ln(1/\mu)$. The
contribution to the renormalization group flow is then
\be
\lim_{\epsilon\to 0}\left\lbrace M\frac{\partial G^{(n)}}{\partial M}\right\rbrace =P^{(n)}+Q^{(n)}+{\cal O}(\mu\ln\mu)~,
\ee
and corresponds to the usual result obtained in $QED$, since we have 
\be
G^{(n)}_{QED}=\left( P^{(n)}+Q^{(n)}\right) \frac{M^\epsilon}{\epsilon}+\mbox{finite}~,
\ee
where the finite terms differ from those in eq.(\ref{finite}).
We stress here that the partial derivative with respect to $M$ 
should be taken at {\it fixed} fermion mass.

\vspace{0.5cm}

\nin{\it Speed of light}\\
If one considers two-loop properties of the model or higher orders, the polarization tensor is affected by Lorentz violation, 
unlike in the
one-loop case, and it is necessary to make sure that the model remains consistent, especially as far as gauge invariance 
and speed of light are concerned.\\
From two loops and above, the field strength gets different corrections for time and space derivatives, and we obtain
\be
 2(1+Y_0)F_{0i}F^{0i}+~(1+Y_1)F_{ij}F^{ij}~,
\ee
where $Y_0$ and $Y_1$ represent the finite quantum corrections to the operators $F_{0i}^2$ and $F_{ij}^2$ respectively,
after absorbing the regularization terms proportional to $1/\epsilon$ or $\ln(1/\mu)$. 
In order to obtain corrections proportional to the Lorentz scalar $F_{\mu\nu}F^{\mu\nu}$, which is necessary to recover gauge 
invariance and the speed of light $c=1$, we rescale the time coordinate and the component $A_0$ as:
\be
t~\to~\frac{t}{\kappa}~~~~~\mbox{and}~~~~~~A_0~\to~\kappa A_0~~~~~\mbox{where}~~~~~~\kappa=\sqrt{\frac{1+Y_1}{1+Y_0}}~.
\ee
One can easily see then, that this rescaling is consistent with gauge invariance of the fermion sector:
\be
\ol\psi\left( i\partial_0-eA_0\right) \gamma^0\psi~\to~\kappa~\ol\psi\left( i\partial_0-eA_0\right) \gamma^0\psi
\ee
The factor $\kappa$, which does not appear in the space components of the covariant derivatives, will then 
contribute to the maximum speed for fermions, together with the corrections to the fermion kinetic terms, 
as explained for the one-loop case. A final identical rescaling for all the gauge field components, by the factor $\sqrt{1+Y_1}$, 
will lead to the redefinition of the coupling constant.

\section{Conclusion}

We have shown that Lorentz-violating higher-order derivatives for an Abelian gauge field lead to a fermion dynamical mass,
independently of the strength of the coupling constant, and that it also leads 
to a consistent quantum theory. The order of magnitude of the fermion dynamical mass is too small for the electron, though, if 
one identifies $M$ with the Plank mass, and this is a problem which could be solved with the introduction of a non-Abelian 
gauge field, as explained below.\\
The main difference with $QED$ is the maximum speed for fermions, which is smaller than
the speed of light. This implies that the fermions see a different effective light cone, but it is not in contradiction with 
the definition of speed of light, since the fermion is always massive.\\
The next step is to extend this study to non-Abelian gauge theories. It is known that vector fields coupled to the axial current 
acquire a mass when a fermion condensate forms \cite{vectormass}, as the result of the massless bound state excitation, which
implies a pole in the polarization tensor for vanishing external momentum. The residue of this pole is the vector mass squared.
In the framework of our present model, the fermion condensate can occur, due to Lorentz-violation
in the Abelian sector, and then generate dynamically the non-Abelian vector masses. 
We note that, in order not to break the non-Abelian
gauge symmetry at the classical level, a Lorentz-violating non-Abelian dynamics 
would involve higher order {\it covariant}-space derivatives, and would therefore introduce new interactions, 
which are not renormalizable. For this reason, we are planning to leave the non-Abelian sector unchanged, and
introduce Lorentz-violation in the Abelian sector only.\\
Another consequence of the coupling of fermions with several gauge fields will be to change the order of 
magnitude of the fermion dynamical mass. Indeed, as was already seen in \cite{hosek}, where a similar study was done,
the contributions of different gauge fields lead to a partial
cancellation in the exponent in the exponential 
of eq.(\ref{mdyn}), giving a phenomenologically more realistic dynamical mass.

\vspace{1cm}

\nin{\bf Acknowledgments} This work is partly supported by the Royal Society, UK.

\section*{Appendix: One-loop Feynman graphs}

We calculate here one-loop corrections to the model (\ref{bare}), assuming that the fermion has mass $m$, which is dynamically generated.
The metric we use throughout is $(+,-,-,-)$.

\vspace{0.5cm}

\nin{\it Fermion kinetic terms}\\
The fermion wave function renormalization is obtained by differentiating the fermion self energy with respect to the external 
momentum $\nu,\vec k$, and then set $\nu=0, \vec k=0$. The fermion self energy is
\bea
&&\Sigma^{(1)}(\nu,\vec k)\\
&=&i(ie)^2\int\frac{d\omega}{2\pi}\int\frac{d^3\vec p}{(2\pi)^3}\frac{\gamma^\mu(\br p+\br k+m)\gamma_\mu}{(1+p^2/M^2)(\omega^2-p^2)
[(\omega+\nu)^2-(\vec p+\vec k)^2-m^2]}\nn
&=&-i\frac{e^2}{(2\pi^4)}\int d\omega\int\frac{d^3\vec p}{1+p^2/M^2}\frac{4m-2(\br p+\br k)}{(\omega^2-p^2)[(\omega+\nu)^2-(\vec p+\vec k)^2-m^2]}
\nonumber~.
\eea
We write 
\be
\Sigma^{(1)}(\nu,\vec k)=Z_0\nu\gamma^0-Z_1\vec k\cdot\vec\gamma+{\cal O}(k^2)~,
\ee
and a derivative with respect to $k^\rho$ gives, after setting $\nu=0, \vec k=0$,
\bea
&&\left.\frac{\partial\Sigma^{(1)}}{\partial k_\rho}\right|_{k=0}\\
&=&\frac{2ie^2}{(2\pi)^4}\int\frac{d^3\vec p}{1+p^2/M^2}\int\frac{d\omega}{\omega^2-p^2}\left(\frac{\gamma^\rho}{\omega^2-p^2-m^2}
-\frac{2p^\rho\br p}{(\omega^2-p^2-m^2)^2}\right) \nonumber
\eea
For $\rho=0$, we obtain, after a Wick rotation,
\be
Z_0=-\frac{e^2}{2\pi^3}\int\frac{p^2dp}{1+p^2/M^2}\int \frac{d\omega}{\omega^2+p^2}
\frac{-\omega^2+p^2+m^2}{(\omega^2+p^2+m^2)^2}
\ee
and integration over the frequency $\omega$ gives 
\bea
Z_0&=&-\frac{e^2}{2\pi^2}\int_0^\infty\frac{x^2dx}{1+\mu^2x^2}\left(2x+\frac{1}{x}-2\sqrt{1+x^2}\right) \nn
&=&-\frac{2\alpha}{\pi}\left(\frac{1}{4}\ln(1/\mu) +\ln2 -\frac{1}{2}\right)+{\cal O}(\mu^2\ln(1/\mu)) ,
\eea
where $\mu=m/M<<1$ and $x=p/m$.\\
For $\rho=i$, we obtain, after a Wick rotation,
\be
Z_1=-\frac{e^2}{2\pi^3}\int\frac{p^2dp}{1+p^2/M^2}\int\frac{d\omega}{\omega^2+p^2}\frac{\omega^2+p^2/3+m^2}{(\omega^2+p^2+m^2)^2}~,
\ee
where the constant 1/3 in factor of $p^2$ in the numerator comes from the Galilean symmetry in the 3-dimensional space. 
The integration over $\omega$ gives
\bea
Z_1&=&-\frac{e^2}{2\pi^2}\int_0^\infty\frac{x^2dx}{1+\mu^2x^2}\left(-\frac{2x}{3}+\frac{1}{x}
+\frac{2x^2/3-1}{\sqrt{1+x^2}}+\frac{x^2/3}{(1+x^2)^{3/2}}\right) \nn
&=&-\frac{2\alpha}{\pi}\left(\frac{1}{4}\ln(1/\mu)+\frac{25}{18}-\frac{5}{3}\ln2\right)+{\cal O}(\mu^2\ln(1/\mu)) ~.
\eea

\vspace{0.5cm}

\nin{\it Vertex}\\
The one-loop correction to the vertex is
\bea
\Gamma_{(1)}^\rho
&=&(ie)^3\int\frac{d\omega}{2\pi}\int\frac{d^3\vec p}{(2\pi)^3}\frac{1}{1+p^2/M^2}
\frac{\gamma^\mu(\br p+m)\gamma^\rho(\br p+m)\gamma_\mu}{(\omega^2-p^2)(\omega^2-p^2-m^2)^2}\nn
&=&-\frac{e^3}{4\pi^3}\int \frac{p^2dp}{1+p^2/M^2}\int d\omega\frac{2\gamma^\rho(\omega^2-p^2)-4p^\rho\br p-2m^2\gamma^\rho}
{(\omega^2-p^2-m^2)^2(\omega^2-p^2)}~.
\eea
For $\rho=0$, we obtain after a Wick rotation
\bea
\Gamma_{(1)}^0&=&-\gamma^0\frac{e^3}{2\pi^3}\int \frac{p^2dp}{1+p^2/M^2}\int d\omega 
\frac{-\omega^2+p^2+m^2}{(\omega^2+p^2)(\omega^2+p^2+m^2)^2}\nn
&=&Z_0e\gamma^0~.
\eea
For $\rho=i$, we obtain after a Wick rotation
\bea
\Gamma_{(1)}^i&=&-\gamma^0\frac{e^3}{2\pi^3}\int \frac{p^2dp}{1+p^2/M^2}\int d\omega 
\frac{\omega^2+p^2/3+m^2}{(\omega^2+p^2)(\omega^2+p^2+m^2)^2}\nn
&=&Z_1e\gamma^i~.
\eea

\end{document}